\documentclass{article}
\usepackage[preprint]{spconf}
\copyrightnotice{979-8-3503-0689-7/23/\$31.00~\copyright2023 IEEE}
\toappear{to Appear in {\it Proc.\ ASRU 2023}}

\usepackage{amsmath,graphicx}

\usepackage{booktabs} %
\usepackage{tabularx}
\usepackage{multirow} %
\usepackage[T1]{fontenc}
\usepackage{setspace}
\usepackage{threeparttable}
\usepackage[labelfont=bf]{caption} %
\usepackage[bookmarks=true,hypertexnames=true,pagebackref=false]{hyperref}
\hypersetup{
  colorlinks=true, %
  urlcolor=black,   %
  linkcolor=blue,  %
  citecolor=red    %
}
\usepackage{color}
\usepackage[table]{xcolor}
\usepackage{cite}
\usepackage{circledsteps}
\usepackage{threeparttable}

\makeatletter
  \newcommand{\linkdest}[1]{\Hy@raisedlink{\hypertarget{#1}{}}}
\makeatother
\newcounter{mathsymbolcount}
\newcommand{\mathsymbol}[1]{%
  \ifcsdef{symbol:#1}{%
    {\hypersetup{hidelinks}\hyperlink{symbol:#1}{#1}}%
  }{%
    \stepcounter{mathsymbolcount}%
    \linkdest{symbol:#1}{}#1%
    \global\expandafter\def\csname symbol:#1\endcsname{}%
  }%
}

\newcolumntype{Y}{>{\centering\arraybackslash}X}

\renewenvironment{thebibliography}[1]{
  \begin{oldthebibliography}{#1}
  \setlength{\itemsep}{-0.1em}
  \setlength{\parskip}{0.em}
}
{
  \end{oldthebibliography}
}

\makeatletter
\def\bstctlcite{\@ifnextchar[{\@bstctlcite}{\@bstctlcite[@auxout]}}
\def\@bstctlcite[#1]#2{\@bsphack
  \@for\@citeb:=#2\do{%
    \edef\@citeb{\expandafter\@firstofone\@citeb}%
    \if@filesw\immediate\write\csname #1\endcsname{\string\citation{\@citeb}}\fi}%
  \@esphack}
\makeatother

\title{Toward Universal Speech Enhancement For Diverse Input Conditions}
\name{Wangyou Zhang$^{1,3}$\thanks{The experiments were done using the PI supercomputer at Shanghai Jiao Tong University and the PSC Bridges2 system via ACCESS allocation CIS210014, supported by National Science Foundation grants \#2138259, \#2138286, \#2138307, \#2137603, and \#2138296.
Wangyou Zhang and Yanmin Qian were supported in part by China STI 2030-Major Projects under Grant No. 2021ZD0201500, in part by China NSFC projects under Grants 62122050 and 62071288, and in part by Shanghai Municipal Science and Technology Major Project under Grant 2021SHZDZX0102.}, Kohei Saijo$^{2,3}$, Zhong-Qiu Wang$^3$, Shinji Watanabe$^3$, Yanmin Qian$^1$}
\address{
    $^1$Shanghai Jiao Tong University, China\;\;
    $^2$Waseda University, Japan\;\;
    $^3$Carnegie Mellon University, USA
}
\begin{document}
\bstctlcite{IEEEexample:BSTcontrol} %
\ninept
\maketitle

\begin{abstract}
The past decade has witnessed substantial growth of data-driven speech enhancement (SE) techniques thanks to deep learning.
While existing approaches have shown impressive performance in some common datasets, most of them are designed only for a single condition (e.g., single-channel, multi-channel, or a fixed sampling frequency) or only consider a single task (e.g., denoising or dereverberation).
Currently, there is no universal SE approach that can effectively handle diverse input conditions with a single model.
In this paper, we make the first attempt to investigate this line of research.
First, we devise a single SE model that is independent of microphone channels, signal lengths, and sampling frequencies.
Second, we design a universal SE benchmark by combining existing public corpora with multiple conditions.
Our experiments on a wide range of datasets show that the proposed single model can successfully handle diverse conditions with strong performance.
\end{abstract}

\begin{figure*}
  \centering
  \linkdest{figure:1}{}
  \includegraphics[width=0.92\textwidth]{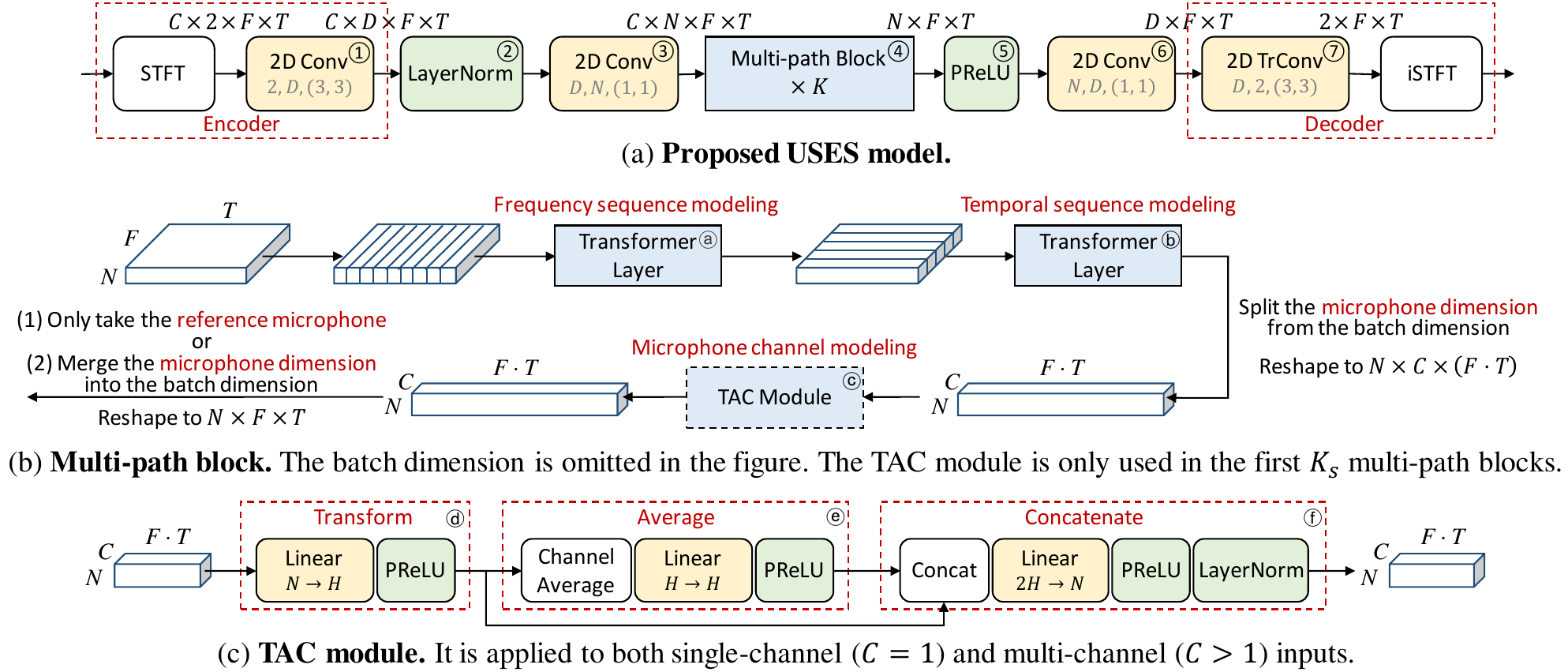}
  \captionsetup{labelfont=bf}
  \caption[overview]{Overview of the proposed versatile SE model. The kernel size and feature maps of convolutional layers are annotated in gray.}
  \label{fig:overview}
  \vspace{-1.8em}
\end{figure*}
\vspace{-1em}
\section{Introduction}
\label{sec:intro}
\vspace{-0.5em}
Speech enhancement (SE) is a task of improving the quality and intelligibility of the speech signal in a noisy and potentially reverberant environment.
Broadly speaking, speech enhancement can be divided into several subtasks such as denoising, dereverberation, echo cancellation, and speech separation~\cite{Springer-Benesty2008}.
The first three mainly focus on single-source conditions, while speech separation tries to separate each speaker's speech from the multi-source mixture recording.
In this paper, we are primarily interested in the first two subtasks.
Therefore, in the remainder of the paper, ``speech enhancement'' refers to denoising and dereverberation.

In recent years, deep learning-based SE techniques have achieved promising performance in various scenarios.
These technique can be roughly classified into three categories: masking-~\cite{Time_frequency-Williamson2017,Glance-Li2022,FRCRN-Zhao2022},  mapping-~\cite{Regression-Xu2014,Complex-Wang2020,Taylor-Li2022,Mask-Liu2023}, and generation-based methods~\cite{SEGAN-Pascual2017,Speech-Maiti2019,MetricGAN_plus-Fu2021,Conditional-Lu2022,Universal-Serra2022,LA_VOcE-Mira2023}.
Masking-based methods estimate a mask either in the time-frequency domain or in the time domain for eliminating noise and reverberation, while mapping-based methods directly estimate the clean-speech representation in the corresponding domain.
Generation-based methods try to reconstruct the clean speech using generation techniques such as generative adversarial networks (GANs)~\cite{SEGAN-Pascual2017,MetricGAN_plus-Fu2021}, diffusion models~\cite{Conditional-Lu2022,Universal-Serra2022}, and resynthesis-based models~\cite{Speech-Maiti2019,LA_VOcE-Mira2023}.
These approaches can provide impressive performance in a condition similar to the training setup.
However, most of the existing approaches are designed only for a single input condition, such as single-channel input, multi-channel input, or input with a fixed sampling frequency.
Recently, there are some attempts to address multiple input conditions with a single model.
For example, the Transform-Average-Concatenate (\mathsymbol{TAC})~\cite{TAC-Luo2020,VarArray-Yoshioka2022} method and a triple-path model~\cite{Time_domain-Pandey22c} are proposed to handle multi-channel signals with a variable number of microphones configured in diverse array geometries.
In~\cite{Continuous-Chen2020}, a continuous speech separation (\mathsymbol{CSS}) approach is proposed to handle arbitrarily-long input with a fixed-length sliding window.
Sampling-frequency-independent models are proposed in~\cite{Sampling_frequency_independent-Saito2021,Sampling-Paulus2022,Efficient-Yu2023} to handle single-channel input with different sampling frequencies.
Nevertheless, these approaches only consider a limited range of input conditions.
To the best of our knowledge, there does not exist a \emph{single-model} approach proposed to handle speech enhancement for single-channel, multi-channel, and arbitrarily long speech signals with different sampling frequencies altogether.

As a step towards universal SE which can handle arbitrary input, in this paper, we aim to devise a single SE model that can handle the aforementioned input conditions without compromising the performance.
We propose an unconstrained speech enhancement and separation network (USES) by carefully integrating several techniques.\footnote{We validate the speech separation and enhancement abilities separately.}
Here, ``unconstrained'' means the model is not constrained to be used only in a fixed input condition.
This single model can accept various forms of input, including 1) single-channel, 2) multi-channel with 3) different array geometries, 4) variable lengths, and 5) variable sampling frequencies.
We also empirically show that the proposed model can be trained on 8 kHz data alone and then tested on data with much higher sampling frequencies (e.g., 48 kHz).
The versatility of this model further inspires us to build a universal SE benchmark to test the performance on various input conditions.
We combine five commonly-used corpora (VoiceBank+DEMAND~\cite{Speech-Valentini-Botinhao2016}, DNS1~\cite{DNS_INTERSPEECH2020-Reddy2020}, CHiME-4~\cite{CHiME4-Vincent2017}, REVERB~\cite{REVERB-Kinoshita2013}, and WHAMR!~\cite{WHAMR-Maciejewski2019}) to train a single SE model that covers a wide range of acoustic scenarios.
The model is then tested on the corresponding test sets with five metrics to comprehensively demonstrate its capability of handling diverse conditions.
Our experiments on various datasets show that the proposed model can successfully cope with different input conditions with strong performance.
The proposed model will be released\footnote{\url{https://github.com/espnet/espnet}} in the ESPnet toolkit~\cite{ESPnet_SE-Li2021}.
We expect this work to attract more attention toward building universal SE models, which can also benefit many downstream speech tasks such as automatic speech recognition (ASR) and speech translation.

\vspace{-1em}
\section{Proposed Model}
\label{sec:model}
\vspace{-0.7em}
In this section, we first describe the overall architecture of the proposed model.
Then, we introduce the key components for handling each of the variable conditions that make our model versatile.

\vspace{-1.2em}
\subsection{Overview}
\label{ssec:overview}
\vspace{-0.5em}
The overall architecture of the proposed model is illustrated in Fig.~\ref{fig:overview}.
We base our proposed approach on a recently proposed dual-path network called time-frequency domain path scanning network (TFPSNet)~\cite{TFPSNet-Yang2022}.
It is one of the top-performing speech separation models in the time-frequency (T-F) domain, and we believe that it can achieve strong performance in speech enhancement as well.
As will be shown in Section~\ref{ssec:sfi}, this model is a natural fit for handling different sampling frequencies.
Without loss of generality, we assume that the input signal contains $\mathsymbol{C}$ microphone channels, where $\mathsymbol{C}$ can be 1 or more.
The encoder consists of a short-time Fourier transform (STFT) module and a subsequent 2D convolutional layer {\hypersetup{hidelinks}\hyperlink{figure:1}{\Circled{\footnotesize 1}}}.
The former converts each input channel into a complex spectrum with shape $2 \times F \times T$, where 2 denotes the real and imaginary parts, $\mathsymbol{F}$ is the number of frequencies, and $\mathsymbol{T}$ the number of frames.
The latter processes each microphone channel independently and projects each T-F bin into a $D$-dimensional embedding for multi-path modeling.
The encoded representations are then processed by channel-wise layer normalization {\hypersetup{hidelinks}\hyperlink{figure:1}{\Circled{\footnotesize 2}}} and projected to a bottleneck dimension $\mathsymbol{N}$ by a point-wise convolutional layer {\hypersetup{hidelinks}\hyperlink{figure:1}{\Circled{\footnotesize 3}}}.
The bottleneck features are processed by $\mathsymbol{K}$ stacked multi-path blocks {\hypersetup{hidelinks}\hyperlink{figure:1}{\Circled{\footnotesize 4}}}, which outputs a single-channel representation of the same shape.
The parametric rectification linear unit (PReLU) activation {\hypersetup{hidelinks}\hyperlink{figure:1}{\Circled{\footnotesize 5}}} is applied to the output, which is later projected back to $\mathsymbol{D}$-dimensional by a point-wise convolutional layer {\hypersetup{hidelinks}\hyperlink{figure:1}{\Circled{\footnotesize 6}}}.
Finally, the output is converted to the complex-valued spectrum via 2D transposed convolution (TrConv, {\hypersetup{hidelinks}\hyperlink{figure:1}{\Circled{\footnotesize 7}}}) and then to waveform via inverse STFT (iSTFT).
We call the proposed method unconstrained speech enhancement and separation (USES) as it can be used in diverse input conditions.\footnote{While we mainly focus on speech enhancement in this paper, we also show in Section~\ref{ssec:exp_separation} that this model works well for speech separation.}

Compared to TFPSNet, we \textbf{make modifications to the encoder and decoder}, following the observations in a recent paper~\cite{TF_GridNet-Wang2023}.
Specifically, we adopt the complex spectral mapping method instead of complex-valued masking in TFPSNet, as it is shown to produce better performance~\cite{TF_GridNet-Wang2023}.
Therefore, the original convolutional layers for mask estimation are replaced with a single 2D convolutional layer {\hypersetup{hidelinks}\hyperlink{figure:1}{\Circled{\footnotesize 6}}}.
The projection layers in the encoder and decoder are also replaced with 2D convolutional {\hypersetup{hidelinks}\hyperlink{figure:1}{\Circled{\footnotesize 1}}} and transposed convolutional {\hypersetup{hidelinks}\hyperlink{figure:1}{\Circled{\footnotesize 7}}} layers, respectively.
The multi-path block {\hypersetup{hidelinks}\hyperlink{figure:1}{\Circled{\footnotesize 4}}} is mostly the same as that in TFPSNet, containing a transformer layer for frequency sequence modeling and another for temporal sequence modeling, as shown in Fig.~\ref{fig:mem} (b).
The transformer layers are the same as those in~\cite{Dual_Path-Chen2020,TFPSNet-Yang2022}.
\textbf{The main differences include} 1) we do not include any T-F path modeling (along the anti-diagonal direction) as we found it not so helpful in the preliminary experiments; 2) we additionally insert a \mathsymbol{TAC} module for channel modeling (Sec~\ref{ssec:tac}).

\begin{figure}[t]
  \centering
  \includegraphics[width=\columnwidth]{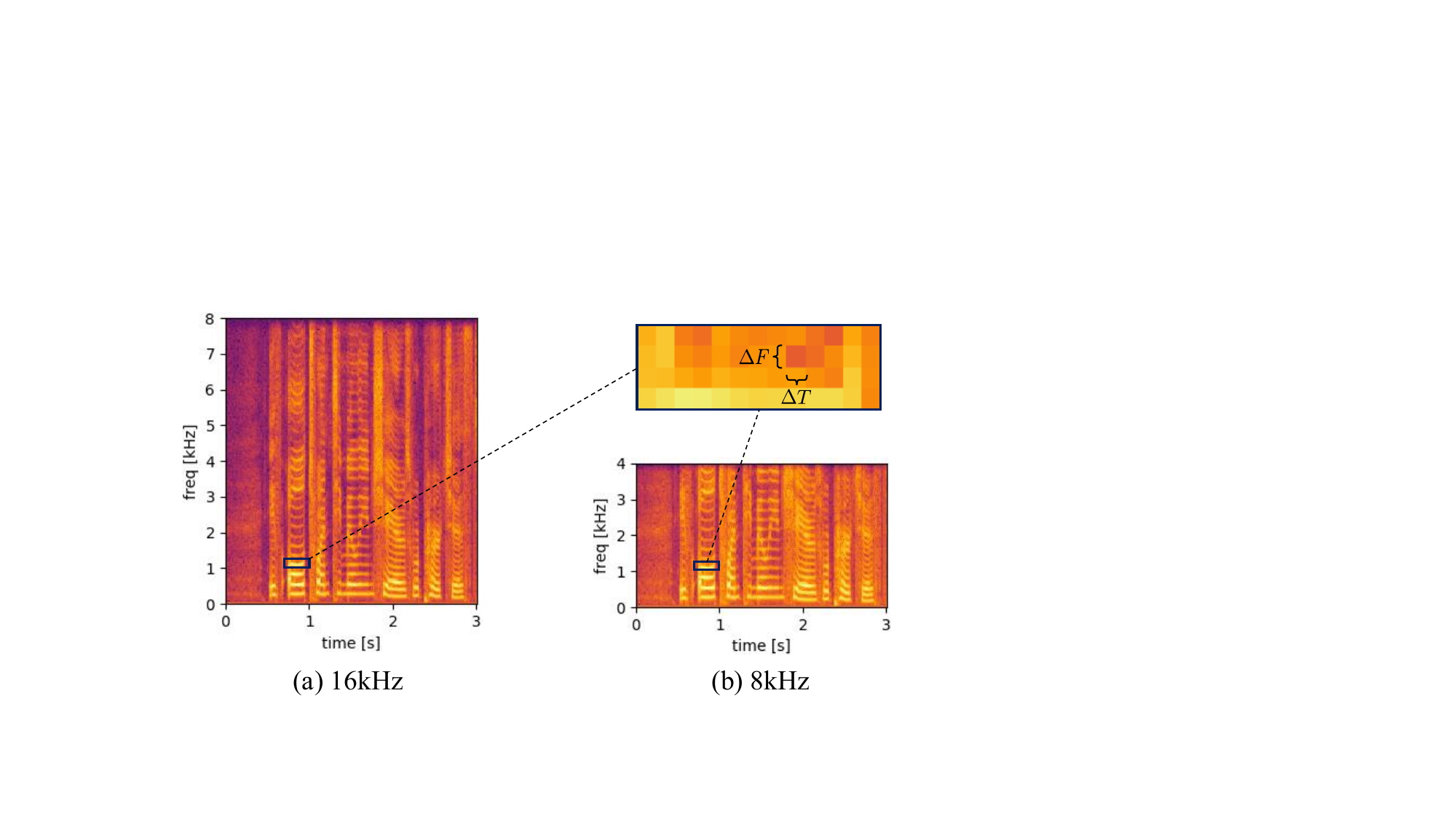}
  \captionsetup{labelfont=bf}
  \caption[sfi]{STFT with fixed-duration window and hop sizes (e.g., 32 ms and 16 ms) will generate spectra with the same frequency and temporal resolution for different sampling frequencies.}
  \label{fig:sfi}
\end{figure}
\vspace{-1em}
\begin{table*}
    \setstretch{0.92}
    \captionsetup{labelfont=bf}
    \caption{Detailed information of the corpora used in our SE experiments. ``\#Ch'' denotes the number of microphone channels in the data. ``T60'' denotes the reverberation time. ``Train.~SNR'' represents signal-to-noise ratio in the training data. ``(Simu)'' and ``(Real)'' denote the synthetic and recorded data, while ``A'' and ``R'' in parentheses represent anechoic and reverberant, respectively.}
    \label{tab:corpora}
    \centering
    \resizebox{1.0\linewidth}{!}{%
        \begin{tabular}{l|ccc|cccc} 
        \toprule
        \textbf{Dataset} & \textbf{Train (hr)} & \textbf{Dev (hr)} & \textbf{Test (hr)} & \textbf{Sampling Freq.} & \textbf{\#Ch} & \textbf{T60 (ms)} & \textbf{Train. SNR (dB)} \\
        \midrule
        \href{https://datashare.ed.ac.uk/handle/10283/2791}{VoiceBank+DEMAND}~\cite{Speech-Valentini-Botinhao2016} & 8.8 & 0.6 & 0.6 & 48kHz & 1 & - & - \\
        \cellcolor[HTML]{EEEEEE} & \cellcolor[HTML]{EEEEEE} & \cellcolor[HTML]{EEEEEE} & \cellcolor[HTML]{EEEEEE}(A) 0.42 & \cellcolor[HTML]{EEEEEE} & \cellcolor[HTML]{EEEEEE} & \cellcolor[HTML]{EEEEEE}(A) -\phantom{300~1300} & \cellcolor[HTML]{EEEEEE} \\
        \cellcolor[HTML]{EEEEEE}\multirow{-2}{*}{\href{https://github.com/microsoft/DNS-Challenge/tree/interspeech2020/master}{DNS1} (v1)~\cite{DNS_INTERSPEECH2020-Reddy2020}} & \cellcolor[HTML]{EEEEEE}\multirow{-2}{*}{(A) 90} & \cellcolor[HTML]{EEEEEE}\multirow{-2}{*}{(A) 10} & \cellcolor[HTML]{EEEEEE}(R) 0.42 & \cellcolor[HTML]{EEEEEE}\multirow{-2}{*}{16kHz} & \cellcolor[HTML]{EEEEEE}\multirow{-2}{*}{1} & \cellcolor[HTML]{EEEEEE}(R) 300{\raise.17ex\hbox{$\scriptstyle\sim$}}1300 & \cellcolor[HTML]{EEEEEE}\multirow{-2}{*}{0{\raise.17ex\hbox{$\scriptstyle\sim$}}40} \\
        \href{https://github.com/microsoft/DNS-Challenge/tree/interspeech2020/master}{DNS1} (v2)~\cite{DNS_INTERSPEECH2020-Reddy2020} & (A) 2700 & (A) 300 & Same as above & 16kHz & 1 & Same as above & -5{\raise.17ex\hbox{$\scriptstyle\sim$}}15 \\
        \cellcolor[HTML]{EEEEEE} & \cellcolor[HTML]{EEEEEE} & \cellcolor[HTML]{EEEEEE} & \cellcolor[HTML]{EEEEEE}(Simu) 2.3 & \cellcolor[HTML]{EEEEEE} & \cellcolor[HTML]{EEEEEE} & \cellcolor[HTML]{EEEEEE} & \cellcolor[HTML]{EEEEEE} \\
        \cellcolor[HTML]{EEEEEE}\multirow{-2}{*}{\href{http://spandh.dcs.shef.ac.uk/chime\_challenge/chime2016/}{CHiME-4}~\cite{CHiME4-Vincent2017}} & \cellcolor[HTML]{EEEEEE}\multirow{-2}{*}{(Simu) 14.7} & \cellcolor[HTML]{EEEEEE}\multirow{-2}{*}{(Simu) 2.9} & \cellcolor[HTML]{EEEEEE}(Real) 2.2 & \cellcolor[HTML]{EEEEEE}\multirow{-2}{*}{16kHz} & \cellcolor[HTML]{EEEEEE}\multirow{-2}{*}{5} & \cellcolor[HTML]{EEEEEE}\multirow{-2}{*}{-} & \cellcolor[HTML]{EEEEEE}\multirow{-2}{*}{{\raise.17ex\hbox{$\scriptstyle\sim$}}5} \\
        \multirow{2}{*}{\href{https://reverb2014.dereverberation.com}{REVERB}~\cite{REVERB-Kinoshita2013}} & \multirow{2}{*}{(Simu) 15.5} & \multirow{2}{*}{(Simu) 3.2} & (Simu) 4.8 & \multirow{2}{*}{16kHz} & \multirow{2}{*}{8} & (Simu) 250, 500, 700 & \multirow{2}{*}{20} \\
        & & & (Real) 0.7 & & & (Real) 700\phantom{250, 500, } & \\
        \cellcolor[HTML]{EEEEEE} & \cellcolor[HTML]{EEEEEE}(A) 58.0 & \cellcolor[HTML]{EEEEEE}(A) 14.7 & \cellcolor[HTML]{EEEEEE}(A) 9.0 & \cellcolor[HTML]{EEEEEE} & \cellcolor[HTML]{EEEEEE} & \cellcolor[HTML]{EEEEEE}(A) -\phantom{100~1000} & \cellcolor[HTML]{EEEEEE} \\
        \cellcolor[HTML]{EEEEEE}\multirow{-2}{*}{\href{https://wham.whisper.ai}{WHAMR!}~\cite{WHAMR-Maciejewski2019}} & \cellcolor[HTML]{EEEEEE}(R) 58.0 & \cellcolor[HTML]{EEEEEE}(R) 14.7 & \cellcolor[HTML]{EEEEEE}(R) 9.0 & \cellcolor[HTML]{EEEEEE}\multirow{-2}{*}{16kHz} & \cellcolor[HTML]{EEEEEE}\multirow{-2}{*}{2} & (\cellcolor[HTML]{EEEEEE}R) 100{\raise.17ex\hbox{$\scriptstyle\sim$}}1000 & \cellcolor[HTML]{EEEEEE}\multirow{-2}{*}{-6{\raise.17ex\hbox{$\scriptstyle\sim$}}3} \\
       \bottomrule
       \end{tabular}%
    }
    \vspace{-1.7em}
\end{table*}
\subsection{Sampling-Frequency-Independent design}
\label{ssec:sfi}
\vspace{-0.5em}
We follow the basic idea in~\cite{Sampling-Paulus2022} for sampling-frequency-independent (\mathsymbol{SFI}) model design.
Namely, we rely on the STFT/iSTFT to obtain consistent T-F representations across different sampling frequencies (\mathsymbol{SF}s).
Since the frequency response of STFT filterbanks shifts linearly for all center frequencies~\cite{Learning-Cornell2022}, it can be easily extended to handle different \mathsymbol{SF}s.
As shown in Fig.~\ref{fig:sfi}, if we use fixed-duration STFT window and hop sizes (e.g., 32 and 16 ms) for different \mathsymbol{SF}s, the resultant spectra will have constant T-F resolution.
As a result, the STFT spectra of the same signal sampled at different \mathsymbol{SF}s will have the same number of frames and different numbers of frequency bins, while the resolution is always consistent.
We can leverage this property to build an \mathsymbol{SFI} model easily as long as the model is capable of handling inputs with two variable dimensions, time and frequency.

Interestingly, the time-frequency domain dual-path models such as TFPSNet\footnote{However, this property is not noticed in the original paper~\cite{TFPSNet-Yang2022}.} and the proposed USES model \emph{inherently satisfy this requirement} and can be directly used for \mathsymbol{SFI} modeling without any modification.
This is because these models treat the SE process as decoupled frequency sequence modeling {\hypersetup{hidelinks}\hyperlink{figure:1}{\Circled{\footnotesize a}}} and temporal sequence modeling {\hypersetup{hidelinks}\hyperlink{figure:1}{\Circled{\footnotesize b}}}, as illustrated in Fig.~\ref{fig:overview} (b), and the transformer layers can naturally process variable-length frequency sequences when different \mathsymbol{SF}s are processed.
In summary, the proposed model is inherently capable of \mathsymbol{SFI} modeling, and all we need is to adaptively adjust the STFT/iSTFT window and hop sizes (to have fixed duration) according to the input \mathsymbol{SF}.

Compared to our method, the \mathsymbol{SFI} convolutional encoder/decoder design in~\cite{Sampling_frequency_independent-Saito2021} is constrained by the maximum frequency range defined in the latent analog filter, which thus limits the highest sampling frequency it can handle\footnote{Our preliminary trial also shows it is less generalizable than STFT.}.
Another recently proposed \mathsymbol{SFI} method in~\cite{Efficient-Yu2023} requires hand-crafted subband division and always resamples the input signal to a pre-defined sampling frequency (e.g., 48 kHz).
Both methods have to be trained with data of different \mathsymbol{SF}s to cover the whole frequency range the model is designed for.
In contrast, our proposed model can be trained with 8 kHz data alone, and then applied to much higher \mathsymbol{SF}s such as 48 kHz.
This also greatly speeds up the training process and reduces the memory consumption during training.

\vspace{-1.3em}
\subsection{Microphone-Channel-Independent design}
\label{ssec:tac}
\vspace{-0.5em}
We adopt the well-developed \mathsymbol{TAC} technique~\cite{TAC-Luo2020} to achieve channel-independent modeling.
As shown in Fig.~\ref{fig:overview} (c), the basic idea of \mathsymbol{TAC} is to project representations of each channel to a hidden dimension $\mathsymbol{H}$ separately {\hypersetup{hidelinks}\hyperlink{figure:1}{\Circled{\footnotesize d}}}, concatenate the channel-averaged representation with each channel's representation {\hypersetup{hidelinks}\hyperlink{figure:1}{\Circled{\footnotesize e}}}, and finally project them back to the original dimension {\hypersetup{hidelinks}\hyperlink{figure:1}{\Circled{\footnotesize f}}}.
The channel number invariance is learned \emph{implicitly} during training.
Similarly to~\cite{VarArray-Yoshioka2022}, we insert the \mathsymbol{TAC} module in the first $\mathsymbol{K_s}$ multi-path blocks for spatial modeling, and then merge the multi-channel representations into single-channel for the rest $(\mathsymbol{K}-\mathsymbol{K_s})$ multi-path blocks.
Instead of averaging the intermediate representations from all channels after the first $\mathsymbol{K_s}$ blocks as in~\cite{VarArray-Yoshioka2022}, we only take the representation at the reference microphone channel and discard the rest.
This is based on the intuition that the information from different channels should be already fused together after the first $\mathsymbol{K_s}$ blocks.
In addition, taking the reference channel allows the model to learn to produce estimates time-aligned with the reference channel, which is often preferable in practice.

\vspace{-1.3em}
\subsection{Signal-Length-Independent design}
\label{ssec:mem}
\begin{figure}[t]
  \centering
  \includegraphics[width=0.98\columnwidth]{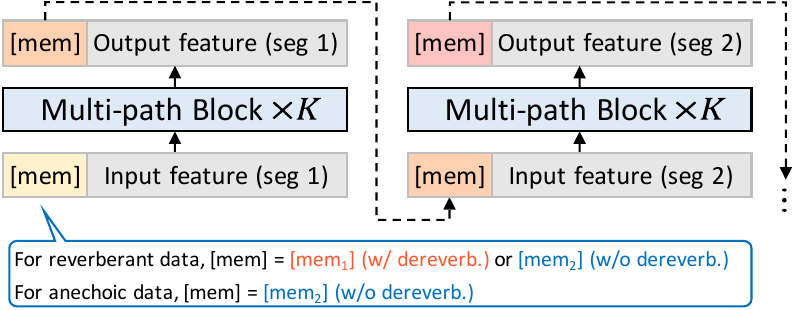}
  \captionsetup{labelfont=bf}
  \caption[mem]{Memory token-based long sequence modeling.}
  \label{fig:mem}
\end{figure}

\vspace{-0.5em}
Inspired by the success of the memory transformer~\cite{Memory-Burtsev2020} in natural language processing for long sequences, we extend the proposed model to handle arbitrarily long input signals following a similar design.
As shown in Fig.~\ref{fig:mem}, we only make a minimal modification to the proposed model by \textbf{adding a group of memory tokens} \texttt{[mem]} of dimension $1 \times N \times 1 \times G$, where $\mathsymbol{G}$ is the group size.
These learnable memory tokens are simply concatenated as a prefix with the feature sequence (output of {\hypersetup{hidelinks}\hyperlink{figure:1}{\Circled{\footnotesize 3}}} in Fig.~\ref{fig:overview}) along the temporal dimension via shape broadcasting.
The concatenated feature is then fed into $\mathsymbol{K}$ multi-path blocks for enhancement, in which the transformer layers could \emph{implicitly} learn to utilize such information via sequence modeling.
The first $\mathsymbol{G}$ frames in the output representation correspond to the processed memory tokens, which are regarded as a summary of the information contained in the current input signal.
These new memory tokens can be then used as the prefix for processing the subsequent input segment.
Thus, we can segment the long-form input into non-overlapping short segments and process each one-by-one without suffering from significant computation and memory costs.
Different from \mathsymbol{CSS}~\cite{Continuous-Chen2020}, we do not need an overlapped sliding window here, as the history information can be retrieved from the output memory tokens from the previous segment.

Furthermore, the learnable memory tokens can be extended to serve as an indicator of different input conditions, similar to the role of prompts in various recent studies~\cite{Prefix_Tuning-Li2021,Exploration-Chang2022}.
To verify this possibility, we design two independent groups of memory tokens (\texttt{[mem}$_1$\texttt{]} and \texttt{[mem}$_2$\texttt{]}) for indicating denoising with and without dereverberation.
As shown in Fig.~\ref{fig:mem}, we apply them accordingly to the reverberant and anechoic data in the extensive SE experiments in Section~\ref{ssec:exp_universal}.

\vspace{-1.2em}
\section{Experiments}
\label{sec:exp}

\vspace{-1.2em}
\subsection{Data}
\label{ssec:data}
\vspace{-0.5em}
\noindent\textbf{Speech separation:} We evaluate the speech separation performance on the commonly-used WSJ0-2mix benchmark~\cite{Deep-Hershey2016} and its spatialized (anechoic) version~\cite{Multi_channel-Wang2018} (\texttt{min} mode).
Each dataset consists of a 30-hour training set, a 10-hour development set, and a 5-hour test set of 2-speaker clean speech mixtures sampled at 8 kHz.
The signal-to-interference ratio (SIR) ranges from -10 to 10 dB.
The spatialized version contains 8 microphone channels, with the microphone arrangement sampled randomly.

\noindent\textbf{Speech enhancement in a single condition:} We train our proposed model on the 16 kHz DNS1 data~\cite{DNS_INTERSPEECH2020-Reddy2020} alone to show its capability in a single condition.
Following existing SE works~\cite{Glance-Li2022,FRCRN-Zhao2022}, we simulate $3000$ hours of non-reverberant data in total, with $2700$ and $300$ hours for training and development, respectively.
The SE performance is then evaluated on the non-blind test set without reverberation, which is around $0.42$ hours.
The detailed information can be found in Table~\ref{tab:corpora} (2nd row).

\noindent\textbf{Speech enhancement in diverse conditions:} To better show the capability of the proposed SE model, we build a comprehensive dataset that can serve as a universal SE benchmark.
The new dataset combines data from five widely-used corpora, as shown in Table~\ref{tab:corpora}, where DNS1 (v2) is not used here to mitigate the data imbalance problem.
The total amount of training data is {\raise.17ex\hbox{$\scriptstyle\sim$}}245 hours.
This dataset covers a wide range of conditions, including single-channel, multi-channel (2ch-8ch), wide-band (16kHz), full-band (48kHz), anechoic, reverberant, and variable-length input in both simulated and real-recorded scenarios.

\vspace{-1.2em}
\subsection{Model and training configurations}
\label{ssec:conf}
\vspace{-0.5em}
In all our experiments, the proposed USES model consists of $\mathsymbol{K}=6$ multi-path blocks, with a \mathsymbol{TAC} module in the first $\mathsymbol{K_s}=3$ blocks for spatial modeling.
The STFT/iSTFT window and hop sizes are always 32 and 16 ms, respectively.
Following TFPSNet~\cite{TFPSNet-Yang2022}, the embedding dimension $\mathsymbol{D}$ is set to $256$ and the bottleneck dimension $\mathsymbol{N}$ to $64$.
The transformer layers in the multi-path blocks have the same configuration as in~\cite{TFPSNet-Yang2022}.
The hidden dimension $\mathsymbol{H}$ in each \mathsymbol{TAC} module is $192$.
When processing multi-channel data, we always take the first channel as the reference channel.
When the memory tokens in Section~\ref{ssec:mem} are applied, we empirically set the number of memory tokens $\mathsymbol{G}$ to $20$, and divide the input signal into non-overlapping segments of {\raise.17ex\hbox{$\scriptstyle\sim$}}1s long (64 frames).
The total number of model parameters is around 3.1 millon.
The pre-trained models and configurations will be released later in ESPnet~\cite{ESPnet_SE-Li2021} for reproducibility.

Our experiments are done based on the ESPnet toolkit~\cite{ESPnet_SE-Li2021}.
The models are trained using the Adam optimizer, and the learning rate increases linearly to \texttt{4e-4} in the first $X$ steps and then decreases by half when the validation performance does not improve for two consecutive epochs.
We set $X$ to 4000 and 25000 for speech separation and enhancement experiments, respectively.
During training, we divide the samples into 4-second chunks to reduce memory costs.
The batch size of all experiments is 4.
We also limit the number of samples for each epoch to 8000.
When training on multi-channel data, we shuffle the channel permutation of each sample and randomly select the number of channels (up to 4 channels) to increase diversity.
We always apply variance normalization to the input signal and revert the variance in the model's output.
For speech separation, all models are trained until convergence (up to 150 epochs) using the SI-SNR loss~\cite{SISDR-LeRoux2019}; and for speech enhancement, we train all models for up to 20 epochs\footnote{For DNS1 data alone, we only train for up to 5 epochs due to the large amount of data, which are enough for the model to converge.} using the loss function proposed in~\cite{Towards-Lu2022}.
The loss function is a scale-invariant multi-resolution $L_1$ loss in the frequency domain plus a time-domain $L_1$ loss term.
We set the STFT window sizes of the multi-resolution $L_1$ loss to \{256, 512, 768, 1024\} and the time-domain loss weight to 0.5.
In each experiment, the model with the best validation performance is selected for evaluation.

We evaluate the SE models with the metrics below:
wide-band PESQ (PESQ-WB)~\cite{PESQ-Rix2001}, STOI~\cite{STOI-Taal2011}, scale-invariant signal-to-noise ratio (SI-SNR)~\cite{SISDR-LeRoux2019}, signal-to-distortion ratio (SDR)~\cite{Performance-Vincent2006}, DNSMOS (OVRL)\footnote{\url{https://github.com/microsoft/DNS-Challenge/blob/master/DNSMOS/DNSMOS/sig_bak_ovr.onnx}}~\cite{DNSMOS_P835-Reddy2022}, and word error rate (WER).
Except for WER, a higher value indicates better performance for all metrics.
The Whisper Large v2 model\footnote{\url{https://huggingface.co/openai/whisper-large-v2}}~\cite{Whisper-Radford2022} is used for WER evaluation.

\begin{table}[t]
    \setstretch{0.92}
    \caption{Speech separation performance on WSJ0-2mix and its spatialized version (\texttt{min} mode). All models are trained only on 8 kHz data, and tested on 8 kHz and 16 kHz data.}
    \label{tab:exp_separation}
    \centering
    \begin{tabular}{lr ccc}
        \toprule
        \textbf{Model} & \textbf{SI-SNRi} & \multicolumn{3}{l}{\hspace{-1em}\textbf{(\,8 kHz} \hspace{1em}/\hspace{3em} \textbf{16 kHz\,)}} \\
        \midrule
        \multicolumn{5}{c}{\textbf{WSJ0-2mix test data}} \\
        \multicolumn{2}{l}{TFPSNet~\cite{TFPSNet-Yang2022}} & \textbf{21.1} & / & - \\
        \multicolumn{2}{l}{TFPSNet (reproduced)} & 21.0 & / & 12.0 (resampling) \\
        \multicolumn{2}{l}{$\ \ $+ \mathsymbol{SFI} STFT/iSTFT} & - & / & 19.7 \\
        \multicolumn{2}{l}{\cellcolor[HTML]{EEEEEE}USES (1ch)} & \cellcolor[HTML]{EEEEEE}20.3 & \cellcolor[HTML]{EEEEEE}/ & \cellcolor[HTML]{EEEEEE}\textbf{19.8} \\
        \multicolumn{2}{l}{\cellcolor[HTML]{EEEEEE}$\ \ $+ mem tokens (1ch)} & \cellcolor[HTML]{EEEEEE}20.9 & \cellcolor[HTML]{EEEEEE}/ & \cellcolor[HTML]{EEEEEE}19.3 \\
        \hline
        \multicolumn{5}{c}{\textbf{1ch spatialized test data}} \\
        \multicolumn{2}{l}{USES (1-2ch)} & 18.9 & / & \textbf{18.4} \\
        \multicolumn{2}{l}{$\ \ $+ mem tokens (1-6ch)} & \textbf{19.9} & / & 18.3 \\
        \hline
        \multicolumn{5}{c}{\textbf{2ch spatialized test data}} \\
        \multicolumn{2}{l}{USES (1-2ch)} & 24.6 & / & 24.2 \\
        \multicolumn{2}{l}{$\ \ $+ mem tokens (1-6ch)} & \textbf{36.1} & / & \textbf{35.0} \\
        \bottomrule
    \end{tabular}%
\end{table}
\vspace{-1em}
\subsection{Evaluation of speech separation performance}
\label{ssec:exp_separation}
\vspace{-0.5em}
We first examine the effectiveness of the three components proposed in Section~\ref{sec:model} by evaluating the speech separation performance on WSJ0-2mix, which makes the comparison with the top-performing TFPSNet~\cite{TFPSNet-Yang2022} convenient.
In addition, the datasets are not large, making it easy to investigate different setups.
We train the model on 8 kHz mixture data, and evaluate the performance on both 8 and 16 kHz test data.
Table~\ref{tab:exp_separation} reports the SI-SNR improvement (SI-SNRi) of models trained on a single channel (denoted as 1ch) and on a variable number of channels (denoted as 1-2ch and 1-6ch).
From the first section in Table~\ref{tab:exp_separation}, we observe that the proposed \mathsymbol{SFI} approach can successfully preserve strong separation performance for the reproduced TFPSNet\footnote{Different from~\cite{TFPSNet-Yang2022}, our reproduction replaces all T-F path scanning transformers with the time-path scanning transformer.} on 16 kHz data.
In comparison, first downsampling 16 kHz mixtures to 8 kHz, then applying TFPSNet trained at 8 kHz for separation, and finally upsampling the separation results to 16 kHz (denoted as ``resampling'' in the second row) suffer from severe SI-SNR degradation.
The proposed USES model also obtains similar performance to TFPSNet on WSJ0-2mix, as our major modifications focus on the invariance to sampling frequencies, microphone channels, and signal lengths.
While applying memory tokens does not change the performance significantly, it enables the model to process variable-length inputs with a constant memory cost during inference.

\setcounter{table}{3}
{
\setlength{\tabcolsep}{3pt}
\begin{table*}
    \caption{Speech enhancement performance of the proposed model in diverse conditions. The models are by default trained only on 8 kHz data, and tested with the original sampling frequencies. ``noisy'' denotes the performance of the input noisy speech, while ``excl'' and ``USES'' denote those of the enhanced speech by corpus-exclusive SE and the proposed single SE models, respectively. ``USES$^+$'' denotes the proposed SE model trained for 5 more epochs (25 epochs in total) with variable \protect\mathsymbol{SF}s (8, 16 and 24 kHz, achieved via downsampling) on the same data. On REVERB and WHAMR! (reverb), the models always do denoising and dereverberation by applying the corresponding memory tokens \texttt{[mem}$_1$\texttt{]}; for other data, only denoising is performed. The best and second best results are made \textbf{bold} and \underline{underlined}, respectively.}    
    \label{tab:exp_universal}
    \centering
\resizebox{\textwidth}{!}{%
\begin{tabular}{lc cccc cccc cccc cccc cccc}
  \toprule
  & &
  \multicolumn{4}{c}{\cellcolor[HTML]{EEEEEE}\textbf{PESQ-WB $\uparrow$}} &
  \multicolumn{4}{c}{\textbf{STOI ($\times100$) $\uparrow$}} &
  \multicolumn{4}{c}{\cellcolor[HTML]{EEEEEE}\textbf{SDR (dB) $\uparrow$}} &
  \multicolumn{4}{c}{\textbf{DNSMOS OVRL $\uparrow$}} &
  \multicolumn{4}{c}{\cellcolor[HTML]{EEEEEE}\textbf{WER (\%) $\downarrow$}} \\
\multirow{-2}{*}{\textbf{Test set}} &
  \multirow{-2}{*}{\textbf{Condition}} &
  \cellcolor[HTML]{EEEEEE}\textbf{noisy} &
  \cellcolor[HTML]{EEEEEE}\textbf{excl} &
  \cellcolor[HTML]{EEEEEE}\textbf{USES} &
  \cellcolor[HTML]{EEEEEE}\textbf{USES$^+$} &
  \textbf{noisy} &
  \textbf{excl} &
  \textbf{USES} &
  \textbf{USES$^+$} &
  \cellcolor[HTML]{EEEEEE}\textbf{noisy} &
  \cellcolor[HTML]{EEEEEE}\textbf{excl} &
  \cellcolor[HTML]{EEEEEE}\textbf{USES} &
  \cellcolor[HTML]{EEEEEE}\textbf{USES$^+$} &
  \textbf{noisy} &
  \textbf{excl} &
  \textbf{USES} &
  \textbf{USES$^+$} &
  \cellcolor[HTML]{EEEEEE}\textbf{noisy} &
  \cellcolor[HTML]{EEEEEE}\textbf{excl} &
  \cellcolor[HTML]{EEEEEE}\textbf{USES} &
  \cellcolor[HTML]{EEEEEE}\textbf{USES$^+$} \\
  \hline
VoiceBank+DEMAND &
  1ch, 48kHz &
  \cellcolor[HTML]{EEEEEE}- &
  \cellcolor[HTML]{EEEEEE}- &
  \cellcolor[HTML]{EEEEEE}- &
  \cellcolor[HTML]{EEEEEE}- &
  92.1 &
  92.8 &
  \underline{93.1} &
  \textbf{95.8} &
  \cellcolor[HTML]{EEEEEE}8.4 &
  \cellcolor[HTML]{EEEEEE}\underline{11.0} &
  \cellcolor[HTML]{EEEEEE}11.0 &
  \cellcolor[HTML]{EEEEEE}\textbf{17.3} &
  2.70 &
  3.10 &
  \underline{3.14} &
  \textbf{3.15} &
  \cellcolor[HTML]{EEEEEE}4.4 &
  \cellcolor[HTML]{EEEEEE}4.2 &
  \cellcolor[HTML]{EEEEEE}\underline{3.2} &
  \cellcolor[HTML]{EEEEEE}\textbf{2.3} \\
VoiceBank+DEMAND &
  1ch, 16kHz &
  \cellcolor[HTML]{EEEEEE}1.98 &
  \cellcolor[HTML]{EEEEEE}\underline{3.08} &
  \cellcolor[HTML]{EEEEEE}\textbf{3.11} &
  \cellcolor[HTML]{EEEEEE}3.06 &
  92.1 &
  \underline{95.3} &
  95.0 &
  \textbf{95.9} &
  \cellcolor[HTML]{EEEEEE}8.5 &
  \cellcolor[HTML]{EEEEEE}20.4 &
  \cellcolor[HTML]{EEEEEE}\underline{21.5} &
  \cellcolor[HTML]{EEEEEE}\textbf{21.8} &
  2.70 &
  3.16 &
  \textbf{3.19} &
  \underline{3.18} &
  \cellcolor[HTML]{EEEEEE}4.4 &
  \cellcolor[HTML]{EEEEEE}\textbf{2.7} &
  \cellcolor[HTML]{EEEEEE}3.1 &
  \cellcolor[HTML]{EEEEEE}\underline{2.8} \\
DNS1 (w/o reverb) &
  1ch, 16kHz &
  \cellcolor[HTML]{EEEEEE}1.58 &
  \cellcolor[HTML]{EEEEEE}3.16 &
  \cellcolor[HTML]{EEEEEE}\underline{3.23} &
  \cellcolor[HTML]{EEEEEE}\textbf{3.35} &
  91.5 &
  97.4 &
  \underline{97.8} &
  \textbf{98.1} &
  \cellcolor[HTML]{EEEEEE}9.1 &
  \cellcolor[HTML]{EEEEEE}\underline{19.9} &
  \cellcolor[HTML]{EEEEEE}19.6 &
  \cellcolor[HTML]{EEEEEE}\textbf{20.5} &
  2.48 &
  3.29 &
  \underline{3.32} &
  \textbf{3.33} &
  \cellcolor[HTML]{EEEEEE}7.2 &
  \cellcolor[HTML]{EEEEEE}\underline{6.2} &
  \cellcolor[HTML]{EEEEEE}6.4 &
  \cellcolor[HTML]{EEEEEE}\textbf{5.9} \\
DNS1 (reverb) &
  1ch, 16kHz &
  \cellcolor[HTML]{EEEEEE}1.82 &
  \cellcolor[HTML]{EEEEEE}1.51 &
  \cellcolor[HTML]{EEEEEE}\underline{2.75} &
  \cellcolor[HTML]{EEEEEE}\textbf{2.92} &
  86.6 &
  68.5 &
  \textbf{89.9} &
  \underline{89.5} &
  \cellcolor[HTML]{EEEEEE}9.2 &
  \cellcolor[HTML]{EEEEEE}2.2 &
  \cellcolor[HTML]{EEEEEE}\underline{13.4} &
  \cellcolor[HTML]{EEEEEE}\textbf{14.0} &
  1.39 &
  2.28 &
  \textbf{2.36} &
  \underline{2.30} &
  \cellcolor[HTML]{EEEEEE}\textbf{19.1} &
  \cellcolor[HTML]{EEEEEE}75.7 &
  \cellcolor[HTML]{EEEEEE}\underline{28.9} &
  \cellcolor[HTML]{EEEEEE}30.4 \\
CHiME-4 (Simu) &
  5ch, 16kHz &
  \cellcolor[HTML]{EEEEEE}1.27 &
  \cellcolor[HTML]{EEEEEE}\textbf{3.16} &
  \cellcolor[HTML]{EEEEEE}\underline{2.95} &
  \cellcolor[HTML]{EEEEEE}\underline{2.95} &
  87.0 &
  \textbf{98.3} &
  97.8 &
  \underline{97.9} &
  \cellcolor[HTML]{EEEEEE}7.54 &
  \cellcolor[HTML]{EEEEEE}\textbf{20.6} &
  \cellcolor[HTML]{EEEEEE}18.3 &
  \cellcolor[HTML]{EEEEEE}\underline{19.1} &
  2.08 &
  \underline{3.22} &
  \underline{3.22} &
  \textbf{3.24} &
  \cellcolor[HTML]{EEEEEE}5.8 &
  \cellcolor[HTML]{EEEEEE}\underline{4.2} &
  \cellcolor[HTML]{EEEEEE}4.4 &
  \cellcolor[HTML]{EEEEEE}\textbf{4.1} \\
REVERB (Simu) &
  8ch, 16kHz &
  \cellcolor[HTML]{EEEEEE}1.48 &
  \cellcolor[HTML]{EEEEEE}1.82 &
  \cellcolor[HTML]{EEEEEE}\textbf{2.09} &
  \cellcolor[HTML]{EEEEEE}\underline{2.08} &
  85.2 &
  88.1 &
  \textbf{89.8} &
  \underline{89.8} &
  \cellcolor[HTML]{EEEEEE}8.7 &
  \cellcolor[HTML]{EEEEEE}10.5 &
  \cellcolor[HTML]{EEEEEE}\underline{11.9} &
  \cellcolor[HTML]{EEEEEE}\textbf{12.2} &
  2.10 &
  \underline{2.84} &
  \textbf{2.98} &
  \textbf{2.98} &
  \cellcolor[HTML]{EEEEEE}4.9 &
  \cellcolor[HTML]{EEEEEE}\underline{4.7} &
  \cellcolor[HTML]{EEEEEE}\textbf{4.6} &
  \cellcolor[HTML]{EEEEEE}\textbf{4.6} \\
WHAMR! (anechoic) &
  2ch, 16kHz &
  \cellcolor[HTML]{EEEEEE}1.11 &
  \cellcolor[HTML]{EEEEEE}\underline{2.62} &
  \cellcolor[HTML]{EEEEEE}2.55 &
  \cellcolor[HTML]{EEEEEE}\textbf{2.69} &
  76.0 &
  \textbf{96.8} &
  96.4 &
  \underline{96.7} &
  \cellcolor[HTML]{EEEEEE}-2.8 &
  \cellcolor[HTML]{EEEEEE}\textbf{16.3} &
  \cellcolor[HTML]{EEEEEE}15.8 &
  \cellcolor[HTML]{EEEEEE}\underline{16.2} &
  1.69 &
  \textbf{3.34} &
  \underline{3.33} &
  \underline{3.33} &
  \cellcolor[HTML]{EEEEEE}8.4 &
  \cellcolor[HTML]{EEEEEE}\textbf{6.0} &
  \cellcolor[HTML]{EEEEEE}\underline{6.2} &
  \cellcolor[HTML]{EEEEEE}\underline{6.2} \\
WHAMR! (reverb) &
  2ch, 16kHz &
  \cellcolor[HTML]{EEEEEE}1.11 &
  \cellcolor[HTML]{EEEEEE}\textbf{2.57} &
  \cellcolor[HTML]{EEEEEE}\underline{2.51} &
  \cellcolor[HTML]{EEEEEE}2.33 &
  73.1 &
  \textbf{96.4} &
  \underline{96.0} &
  94.7 &
  \cellcolor[HTML]{EEEEEE}-1.8 &
  \cellcolor[HTML]{EEEEEE}\textbf{14.6} &
  \cellcolor[HTML]{EEEEEE}\underline{13.8} &
  \cellcolor[HTML]{EEEEEE}12.9 &
  1.41 &
  \textbf{3.32} &
  \textbf{3.32} &
  \underline{3.28} &
  \cellcolor[HTML]{EEEEEE}9.3 &
  \cellcolor[HTML]{EEEEEE}\textbf{6.5} &
  \cellcolor[HTML]{EEEEEE}\textbf{6.5} &
  \cellcolor[HTML]{EEEEEE}\underline{6.9} \\
  \bottomrule
\end{tabular}%
}
\end{table*}
\setcounter{table}{2}
\begin{table}[th]
    \setstretch{0.95}
    \vspace{-1.5em}
    \caption{Speech enhancement performance on DNS1 non-blind test set (without reverberation). All models are trained on 16 kHz data. Our models are non-causal.}
    \label{tab:exp_enhancement}
    \centering
    \resizebox{\columnwidth}{!}{%
    \begin{tabular}{lccc}
        \toprule
        \textbf{Model} & \textbf{PESQ-WB $\uparrow$} & \textbf{STOI ($\times100$) $\uparrow$} & \textbf{SI-SNR (dB) $\uparrow$} \\
        \midrule
        Noisy & 1.58 & 91.5 & 9.07 \\
        GaGNet~\cite{Glance-Li2022} (causal) & 3.17 & 97.1 & 18.9 \\
        FRCRN~\cite{FRCRN-Zhao2022} (causal) & 3.23 & 97.7 & 19.8 \\
        MFNet~\cite{Mask-Liu2023} (non-causal) & 3.43 & 98.0 & 20.3 \\
        \hline
        Proposed (DNS1 v1) & 3.16 & 97.4 & 19.9  \\
        Proposed (DNS1 v2) & \textbf{3.46} & \textbf{98.1} & \textbf{21.2} \\
        \bottomrule
    \end{tabular}%
    }
\end{table}
}
\setcounter{table}{4}

For experiments on the spatialized WSJ0-2mix data, we can see that the model achieves very similar performance on both 8 and 16 kHz data.
Although the single-channel SI-SNR performance degrades {\raise.17ex\hbox{$\scriptstyle\sim$}}1.4 dB, the multi-channel performance becomes much better, even with only 2 input channels.
Further applying the memory tokens and increasing the input channels during training can improve the performance, especially for the multi-channel case.

\vspace{-1em}
\subsection{Evaluation of SE performance in a single condition}
\label{ssec:exp_enhancement}
\vspace{-0.5em}
Given the success of the universal properties of USES in a controlled experimental condition in Section~\ref{ssec:exp_separation}, this subsection compares the SE performance of the proposed model with memory tokens to existing methods on the DNS1 non-blind test data.
Our models are respectively trained on the simulated DNS1 v1 and v2 data without reverberation, as described in Table~\ref{tab:corpora}.
Due to the large amount of training data and limited time, we only train the proposed model for {\raise.17ex\hbox{$\scriptstyle\sim$}}0.5 epochs on DNS1 v2 data, covering around 45\% of the entire training set.
For DNS1 v1 data, we train the model for 5 epochs, which is around 3.7 passes of the entire training set\footnote{The training of both models is well converged with our setup.}.
The SE performance is presented in Table~\ref{tab:exp_enhancement}, where we compare our proposed model with the top-performing methods on DNS1 data.
We can see that although our models are only trained for very limited steps, they can still achieve very competitive performance compared to the existing SE methods.
The model trained on DNS1 v2 data achieves a new state-of-the-art performance while only trained for {\raise.17ex\hbox{$\scriptstyle\sim$}}0.5 epochs.
However, it should be noted that our proposed model and MFNet~\cite{Mask-Liu2023} are non-causal, while the other listed models are causal.
Therefore, we cannot make a fair comparison here.
Nevertheless, this result at least demonstrates the effectiveness of the proposed method in the standard SE task.

\vspace{-1em}
\subsection{Evaluation of SE performance in diverse conditions}
\label{ssec:exp_universal}
\vspace{-0.5em}
Finally, we present the SE performance across different conditions.
Here, we adopt the same architecture as in Section~\ref{ssec:exp_enhancement} as the base model, and train two groups of memory tokens as mentioned in Section~\ref{ssec:mem} to control whether dereverberation is applied or not.
During training, we always resample the data to 8 kHz to reduce memory costs, and evaluate the performance on the original data.
For the VoiceBank+DEMAND 48 kHz test data, the original and enhanced audios are resampled to 16 kHz before evaluating STOI, DNSMOS, and WER.
We can see in Table~\ref{tab:exp_universal} that, in all conditions, the proposed single SE model (\textbf{USES} columns) achieves strong enhancement performance that is on par with or better than the corresponding corpus-exclusive SE model (\textbf{excl} columns).
Note that the DNS1 (reverb) test data represents an unseen noisy-reverberant condition during training.
In this condition, the proposed SE model achieves much better performance, which shows the benefit of training on diverse data using the proposed model.
However, we can see that, on the 48 kHz VoiceBank+DEMAND test set, both SE models suffer from SDR degradation.
Our investigation implies that it is caused by the greatly increased frequency bins in 48 kHz (129$\rightarrow$769), as the model is only trained on 8 kHz.
To mitigate this mismatch, we continue training the proposed SE model for 5 more epochs (\textbf{USES$^+$} columns) with variable \mathsymbol{SF}s on the same data.
We can see that increasing the \mathsymbol{SF} diversity consistently improves the SE performance in different conditions, which shows the capacity of our model.
The enhanced audios are available on our demo page: {\footnotesize\url{https://Emrys365.github.io/Universal-SE-demo/}}.

For ASR performance evaluation on different datasets\footnote{For DNS1 test data, we prepare the transcription  manually as the reference, which is available at \href{https://github.com/Emrys365/DNS_text}{\texttt{github.com/Emrys365/DNS\_text}}.}, we use the same Whisper Large model without external language models.
The same text normalization~\cite{Whisper-Radford2022} is applied to both reference transcripts and Whisper outputs.
As shown in Table~\ref{tab:exp_universal}, the proposed SE model achieves similar ASR performance to the corpus-exclusive SE model in most conditions.
We further evaluate the performance on two real-recorded datasets in Table~\ref{tab:exp_real}.
On the REVERB (Real) data, the SE models perform well in terms of both enhancement and ASR.
On the CHiME-4 (Real) data, we notice that some microphone channels contain much noisier signals, which cannot be well processed by the \mathsymbol{TAC} module inherently, as it simply averages all channels for fusion.
Therefore, we only conduct single-channel SE on the reference channel (CH5) in this case.
The proposed SE model achieves much better performance than the corpus-exclusive model, coinciding with the observation in Table~\ref{tab:exp_universal}.
Note that on DNS1 (reverb) and CHiME-4 (Real), the SE model does not improve the ASR performance, which is a commonly observed phenomenon~\cite{Closing-Zhang2021,Learning-Sato2022,How-Iwamoto2022} due to the introduced artifacts by enhancement in mismatched conditions.

\begin{table}[t]
    \vspace{-1.5em}
    \caption{DNSMOS and ASR results on real-recorded data.}
    \label{tab:exp_real}
    \centering
    \resizebox{\columnwidth}{!}{%
    \setlength{\tabcolsep}{2pt}
    \begin{threeparttable}
    \begin{tabular}{l cccc cccc}
        \toprule
        \multirow{2}{*}{\textbf{Test set}} & \multicolumn{4}{c}{\cellcolor[HTML]{EEEEEE}\textbf{DNSMOS OVRL} $\uparrow$} & \multicolumn{4}{c}{\textbf{WER (\%)} $\downarrow$} \\
        & \cellcolor[HTML]{EEEEEE}\textbf{noisy} & \cellcolor[HTML]{EEEEEE}\textbf{excl} & \cellcolor[HTML]{EEEEEE}\textbf{USES} & \cellcolor[HTML]{EEEEEE}\textbf{USES$^+$} & \textbf{noisy} & \textbf{excl} & \textbf{USES} & \textbf{USES$^+$} \\
        \hline
        CHiME-4 (Real)$^*$ & \cellcolor[HTML]{EEEEEE}1.46 & \cellcolor[HTML]{EEEEEE}2.94 & \cellcolor[HTML]{EEEEEE}\underline{3.07} & \cellcolor[HTML]{EEEEEE}\textbf{3.12} & \textbf{6.7} & 11.0 & 7.4 & \underline{7.1} \\
        REVERB (Real) & \cellcolor[HTML]{EEEEEE}1.57 & \cellcolor[HTML]{EEEEEE}2.25 & \cellcolor[HTML]{EEEEEE}\textbf{3.11} & \cellcolor[HTML]{EEEEEE}\underline{3.07} & 5.8 & 5.4 & \underline{5.1} & \textbf{5.0} \\
        \bottomrule
    \end{tabular}%
    \begin{tablenotes}[flushleft]\footnotesize
        \item[*] Single-channel SE on CH5 is used in CHiME-4 (Real). (See Section~\ref{ssec:exp_universal})
    \end{tablenotes}
    \end{threeparttable}%
    }
\end{table}

\vspace{-1.2em}
\section{Conclusion}
\label{sec:conclusion}
\vspace{-0.5em}
In this paper, we have devised a single speech enhancement model USES that can handle denoising and dereverberation in diverse input conditions altogether, including variable microphone channels, sampling frequencies, signal lengths, and different environments.
Experiments on a wide range of datasets show that the proposed model can achieve very competitive performance for both speech separation and speech enhancement tasks.
We further design a benchmark for evaluating the universal SE performance across various conditions, which also reveals some less-explored aspects in the SE literature such as the generalizability across different domains.
We hope this contribution can attract more efforts toward building universal SE models for real-world speech applications.

\bibliographystyle{IEEEtran}
\bibliography{refs}

\begin{thebibliography}{10}
\providecommand{\url}[1]{#1}
\csname url@samestyle\endcsname
\providecommand{\newblock}{\relax}
\providecommand{\bibinfo}[2]{#2}
\providecommand{\BIBentrySTDinterwordspacing}{\spaceskip=0pt\relax}
\providecommand{\BIBentryALTinterwordstretchfactor}{4}
\providecommand{\BIBentryALTinterwordspacing}{\spaceskip=\fontdimen2\font plus
\BIBentryALTinterwordstretchfactor\fontdimen3\font minus
  \fontdimen4\font\relax}
\providecommand{\BIBforeignlanguage}[2]{{%
\expandafter\ifx\csname l@#1\endcsname\relax
\typeout{** WARNING: IEEEtran.bst: No hyphenation pattern has been}%
\typeout{** loaded for the language `#1'. Using the pattern for}%
\typeout{** the default language instead.}%
\else
\language=\csname l@#1\endcsname
\fi
#2}}
\providecommand{\BIBdecl}{\relax}
\BIBdecl

\bibitem{Springer-Benesty2008}
J.~Benesty \emph{et~al.}, \emph{Springer handbook of speech processing}.\hskip
  1em plus 0.5em minus 0.4em\relax Springer, 2008, vol.~1.

\bibitem{Time_frequency-Williamson2017}
D.~S. Williamson \emph{et~al.}, ``Time-frequency masking in the complex domain
  for speech dereverberation and denoising,'' \emph{IEEE/ACM Trans. ASLP.},
  vol.~25, no.~7, pp. 1492--1501, 2017.

\bibitem{Glance-Li2022}
A.~Li \emph{et~al.}, ``Glance and gaze: A collaborative learning framework for
  single-channel speech enhancement,'' \emph{Applied Acoustics}, vol. 187, p.
  108499, 2022.

\bibitem{FRCRN-Zhao2022}
S.~Zhao \emph{et~al.}, ``{FRCRN}: Boosting feature representation using
  frequency recurrence for monaural speech enhancement,'' in \emph{ICASSP},
  2022, pp. 9281--9285.

\bibitem{Regression-Xu2014}
Y.~Xu \emph{et~al.}, ``A regression approach to speech enhancement based on
  deep neural networks,'' \emph{IEEE/ACM Trans. ASLP.}, vol.~23, no.~1, pp.
  7--19, 2014.

\bibitem{Complex-Wang2020}
Z.-Q. Wang \emph{et~al.}, ``Complex spectral mapping for single-and
  multi-channel speech enhancement and robust {ASR},'' \emph{IEEE/ACM Trans.
  ASLP.}, vol.~28, pp. 1778--1787, 2020.

\bibitem{Taylor-Li2022}
A.~Li \emph{et~al.}, ``{Taylor}, can you hear me now? a {Taylor}-unfolding
  framework for monaural speech enhancement,'' in \emph{Proc. IJCAI}, 2022, pp.
  4193--4200.

\bibitem{Mask-Liu2023}
L.~Liu \emph{et~al.}, ``A mask free neural network for monaural speech
  enhancement,'' in \emph{Interspeech}, 2023, pp. 2468--2472.

\bibitem{SEGAN-Pascual2017}
S.~Pascual \emph{et~al.}, ``{SEGAN}: Speech enhancement generative adversarial
  network,'' in \emph{Interspeech}, 2017, pp. 3642--3646.

\bibitem{Speech-Maiti2019}
S.~Maiti and M.~I. Mandel, ``Speech denoising by parametric resynthesis,'' in
  \emph{ICASSP}, 2019, pp. 6995--6999.

\bibitem{MetricGAN_plus-Fu2021}
S.-W. Fu \emph{et~al.}, ``{MetricGAN+}: An improved version of {MetricGAN} for
  speech enhancement,'' in \emph{Interspeech}, 2021, pp. 201--205.

\bibitem{Conditional-Lu2022}
Y.-J. Lu \emph{et~al.}, ``Conditional diffusion probabilistic model for speech
  enhancement,'' in \emph{ICASSP}, 2022, pp. 7402--7406.

\bibitem{Universal-Serra2022}
J.~Serr{\`a} \emph{et~al.}, ``Universal speech enhancement with score-based
  diffusion,'' \emph{arXiv preprint arXiv:2206.03065}, 2022.

\bibitem{LA_VOcE-Mira2023}
R.~Mira \emph{et~al.}, ``{LA-VocE}: Low-{SNR} audio-visual speech enhancement
  using neural vocoders,'' in \emph{ICASSP}, 2023, pp. 1--5.

\bibitem{TAC-Luo2020}
Y.~Luo \emph{et~al.}, ``End-to-end microphone permutation and number invariant
  multi-channel speech separation,'' in \emph{ICASSP}, 2020, pp. 6394--6398.

\bibitem{VarArray-Yoshioka2022}
T.~Yoshioka \emph{et~al.}, ``{VarArray}: Array-geometry-agnostic continuous
  speech separation,'' in \emph{ICASSP}, 2022, pp. 6027--6031.

\bibitem{Time_domain-Pandey22c}
A.~Pandey \emph{et~al.}, ``Time-domain ad-hoc array speech enhancement using a
  triple-path network,'' in \emph{Interspeech}, 2022, pp. 729--733.

\bibitem{Continuous-Chen2020}
Z.~Chen \emph{et~al.}, ``Continuous speech separation: Dataset and analysis,''
  in \emph{ICASSP}, 2020, pp. 7284--7288.

\bibitem{Sampling_frequency_independent-Saito2021}
K.~Saito \emph{et~al.}, ``Sampling-frequency-independent audio source
  separation using convolution layer based on impulse invariant method,'' in
  \emph{Proc. EUSIPCO}, 2021, pp. 321--325.

\bibitem{Sampling-Paulus2022}
J.~Paulus and M.~Torcoli, ``Sampling frequency independent dialogue
  separation,'' in \emph{Proc. EUSIPCO}, 2022, pp. 160--164.

\bibitem{Efficient-Yu2023}
J.~Yu and Y.~Luo, ``Efficient monaural speech enhancement with universal sample
  rate band-split {RNN},'' in \emph{ICASSP}, 2023, pp. 1--5.

\bibitem{Speech-Valentini-Botinhao2016}
C.~Valentini-Botinhao \emph{et~al.}, ``Speech enhancement for a noise-robust
  text-to-speech synthesis system using deep recurrent neural networks,'' in
  \emph{Interspeech}, 2016, pp. 352--356.

\bibitem{DNS_INTERSPEECH2020-Reddy2020}
C.~K. Reddy \emph{et~al.}, ``The {INTERSPEECH} 2020 deep noise suppression
  challenge: Datasets, subjective testing framework, and challenge results,''
  in \emph{Interspeech}, 2020, pp. 2492--2496.

\bibitem{CHiME4-Vincent2017}
E.~Vincent \emph{et~al.}, ``An analysis of environment, microphone and data
  simulation mismatches in robust speech recognition,'' \emph{Computer Speech
  \& Language}, vol.~46, pp. 535--557, 2017.

\bibitem{REVERB-Kinoshita2013}
K.~Kinoshita \emph{et~al.}, ``The {REVERB} challenge: A common evaluation
  framework for dereverberation and recognition of reverberant speech,'' in
  \emph{Proc. IEEE WASPAA}, 2013, pp. 1--4.

\bibitem{WHAMR-Maciejewski2019}
M.~Maciejewski \emph{et~al.}, ``{WHAMR!}: Noisy and reverberant single-channel
  speech separation,'' in \emph{ICASSP}, 2019, pp. 696--700.

\bibitem{ESPnet_SE-Li2021}
C.~Li \emph{et~al.}, ``{ESPnet-SE}: End-to-end speech enhancement and
  separation toolkit designed for {ASR} integration,'' in \emph{Proc. IEEE
  SLT}, 2021, pp. 785--792.

\bibitem{TFPSNet-Yang2022}
L.~Yang \emph{et~al.}, ``{TFPSNet}: Time-frequency domain path scanning network
  for speech separation,'' in \emph{ICASSP}, 2022, pp. 6842--6846.

\bibitem{TF_GridNet-Wang2023}
Z.-Q. Wang \emph{et~al.}, ``{TF-GridNet}: Making time-frequency domain models
  great again for monaural speaker separation,'' in \emph{ICASSP}, 2023, pp.
  1--5.

\bibitem{Dual_Path-Chen2020}
J.~Chen \emph{et~al.}, ``Dual-path transformer network: Direct context-aware
  modeling for end-to-end monaural speech separation,'' in \emph{Interspeech},
  2020, pp. 2642--2646.

\bibitem{Learning-Cornell2022}
S.~Cornell \emph{et~al.}, ``Learning filterbanks for end-to-end acoustic
  beamforming,'' in \emph{ICASSP}, 2022, pp. 6507--6511.

\bibitem{Memory-Burtsev2020}
M.~S. Burtsev \emph{et~al.}, ``Memory transformer,'' \emph{arXiv preprint
  arXiv:2006.11527}, 2020.

\bibitem{Prefix_Tuning-Li2021}
X.~L. Li and P.~Liang, ``{Prefix-Tuning}: Optimizing continuous prompts for
  generation,'' in \emph{Proc. ACL/IJCNLP}, 2021, pp. 4582--4597.

\bibitem{Exploration-Chang2022}
K.-W. Chang \emph{et~al.}, ``An exploration of prompt tuning on generative
  spoken language model for speech processing tasks,'' in \emph{Interspeech},
  2022, pp. 5005--5009.

\bibitem{Deep-Hershey2016}
J.~R. Hershey \emph{et~al.}, ``Deep clustering: Discriminative embeddings for
  segmentation and separation,'' in \emph{ICASSP}, 2016, pp. 31--35.

\bibitem{Multi_channel-Wang2018}
Z.-Q. Wang \emph{et~al.}, ``Multi-channel deep clustering: Discriminative
  spectral and spatial embeddings for speaker-independent speech separation,''
  in \emph{ICASSP}, 2018, pp. 1--5.

\bibitem{SISDR-LeRoux2019}
J.~Le~Roux \emph{et~al.}, ``{SDR}---half-baked or well done?'' in
  \emph{ICASSP}, 2019, pp. 626--630.

\bibitem{Towards-Lu2022}
Y.-J. Lu \emph{et~al.}, ``Towards low-distortion multi-channel speech
  enhancement: The {ESPnet}-{SE} submission to the {L3DAS22} challenge,'' in
  \emph{ICASSP}, 2022, pp. 9201--9205.

\bibitem{PESQ-Rix2001}
A.~W. Rix \emph{et~al.}, ``Perceptual evaluation of speech quality ({PESQ})---a
  new method for speech quality assessment of telephone networks and codecs,''
  in \emph{ICASSP}, vol.~2, 2001, pp. 749--752.

\bibitem{STOI-Taal2011}
C.~H. Taal \emph{et~al.}, ``An algorithm for intelligibility prediction of
  time--frequency weighted noisy speech,'' \emph{IEEE Trans. ASLP.}, vol.~19,
  no.~7, pp. 2125--2136, 2011.

\bibitem{Performance-Vincent2006}
E.~Vincent \emph{et~al.}, ``Performance measurement in blind audio source
  separation,'' \emph{IEEE Trans. ASLP.}, vol.~14, no.~4, pp. 1462--1469, 2006.

\bibitem{DNSMOS_P835-Reddy2022}
C.~K. Reddy \emph{et~al.}, ``{DNSMOS} {P}.835: A non-intrusive perceptual
  objective speech quality metric to evaluate noise suppressors,'' in
  \emph{ICASSP}, 2022, pp. 886--890.

\bibitem{Whisper-Radford2022}
A.~Radford \emph{et~al.}, ``Robust speech recognition via large-scale weak
  supervision,'' \emph{arXiv preprint arXiv:2212.04356}, 2022.

\bibitem{Closing-Zhang2021}
W.~Zhang \emph{et~al.}, ``Closing the gap between time-domain multi-channel
  speech enhancement on real and simulation conditions,'' in \emph{Proc. IEEE
  WASPAA}, 2021, pp. 146--150.

\bibitem{Learning-Sato2022}
H.~Sato \emph{et~al.}, ``Learning to enhance or not: Neural network-based
  switching of enhanced and observed signals for overlapping speech
  recognition,'' in \emph{ICASSP}, 2022, pp. 6287--6291.

\bibitem{How-Iwamoto2022}
K.~Iwamoto \emph{et~al.}, ``How bad are artifacts?: Analyzing the impact of
  speech enhancement errors on {ASR},'' in \emph{Interspeech}, 2022, pp.
  5418--5422.

\end{thebibliography}

\end{document}